\documentclass[fleqn,twoside]{article}
\usepackage{espcrc2}

\usepackage{graphicx}

\newcommand{\AmS}{{\protect\the\textfont2
  A\kern-.1667em\lower.5ex\hbox{M}\kern-.125emS}}

\title{Staggered Fermion, its Symmetry and Ichimatsu-Patterned Lattice}
\author{K. Itoh\address{Faculty of Education, 
        Niigata University, Niigata 950-2181, Japan},
        M. Kato\address{Institute of Physics, University of Tokyo, Komaba, 
                        Meguroku, Tokyo 153-8902, Japan},
        M. Murata\address[GSNU]{Graduate School of Natural Science and Technology, 
Niigata University, 
        Niigata 950-2181, Japan},
        H. Sawanaka\addressmark[GSNU]
        and
        H. So\address{Department of Physics, Niigata University, 
        Niigata 950-2181, Japan}\thanks{Talk presented by H. So. This work was 
        supported in part by Grants-in-Aid for Scientific Research No. 13135209 
 from the Japan Society for the Promotion of Science.}
}       
       
\begin{document}

\begin{abstract}
We investigate exact symmetries of a staggered fermion in D dimensions. 
The Dirac operator is reformulated  by SO(2D) Clifford algebra. 
The chiral symmetry, rotational 
invariance and parity symmetries are clarified  in any dimension. 
Local scalar and pseudo-scalar modes are definitely  determined, in which we find 
non-standard  modes. The relation to Ichimatsu-patterned lattice approach   
is discussed. 
\vspace{1pc}
\end{abstract}

\maketitle

\section{INTRODUCTION}
When we consider super Yang-Mills theory on a lattice, 
gauge fields are put on links and fermi fields are put on sites on the staggered way
\cite{Itoh:2001rx,Itoh:2002nq}. 
The  staggered fermion is formulated for the purpose of solving 
doubling phenomena in lattice fermion\cite{Kogut:1974ag,Susskind:1976jm}. Although the number of doubling is translated 
into that of flavor, it is not clear whether each flavor 
is a  spinor or not.  Discrete space-time symmetries such as P, C 
of the fermion  have  not  been defined.  
Specially, for the chiral symmetry, nobody answers why there exist a mass-protect symmetry 
 for odd dimensions and what is the symmetry. 

In this talk, we present the new formulation of a staggered fermion 
based on cell's idea in our Ichimatsu-patterned lattice and 
SO(2D) Clifford algebra in a D-dimensional lattice.  
The transformation of a staggered fermion 
for the space-time rotation  is clarified.  
The associated  symmetries(chiral and parity) are also investigated 
for the staggered fermion. 
The bi-linear operators for scalar and pseudo-scalar modes are found 
owing to the rotational transformation.  Further details will be presented in \cite{IKMSS}.

\section{FORMULATION}

A staggered fermion, $\xi_n$, has been formulated by putting  
a Grassmann variable on a site, $n$.  For  fermion, we need to obtain the spinorization 
for the staggered Grassmann variable.  Usually, a sign factor,  
\begin{equation}
\eta_{\mu}(n) = (-1)^{\sum_{\nu < \mu} n_{\nu}} ,
\end{equation}
appears in  the staggered fermion\cite{Kawamoto:1981hw,Kluberg-Stern:1983dg}. But the validity is held only when 
the spinor has $2^{\frac{D}{2}}$-components on a D-dimensional lattice.   
It must be shown that single component staggered fermion is reconstructed as 
the Dirac spinor.  

The factor is defined on a link, $(n,\mu)$ 
on the lattice and has a property of modulo-2 translational invariance, 

\begin{equation}
\eta_{\mu}(n+2\hat{\nu}) = \eta_{\mu}(n)  .
\end{equation}
This means a just Ichimatsu-pattern\cite{Itoh:2001rx,Itoh:2002nq}. See Fig.~1.   
In the pattern, a minimal unit is  so-called a cell, 
i.e. a fundamental lattice and its coordinates, $(N,r)$,  are expressed as
 
\begin{equation}
n_{\mu} = r_{\mu} + 2N_{\mu}  .
\end{equation}
where $r_{\mu}$ has only  value of 0 or 1 and $N_{\mu}$ implies the position of the cell with 
double lattice constant $2a$. 
  
\begin{center}
\includegraphics[width=5.0cm]{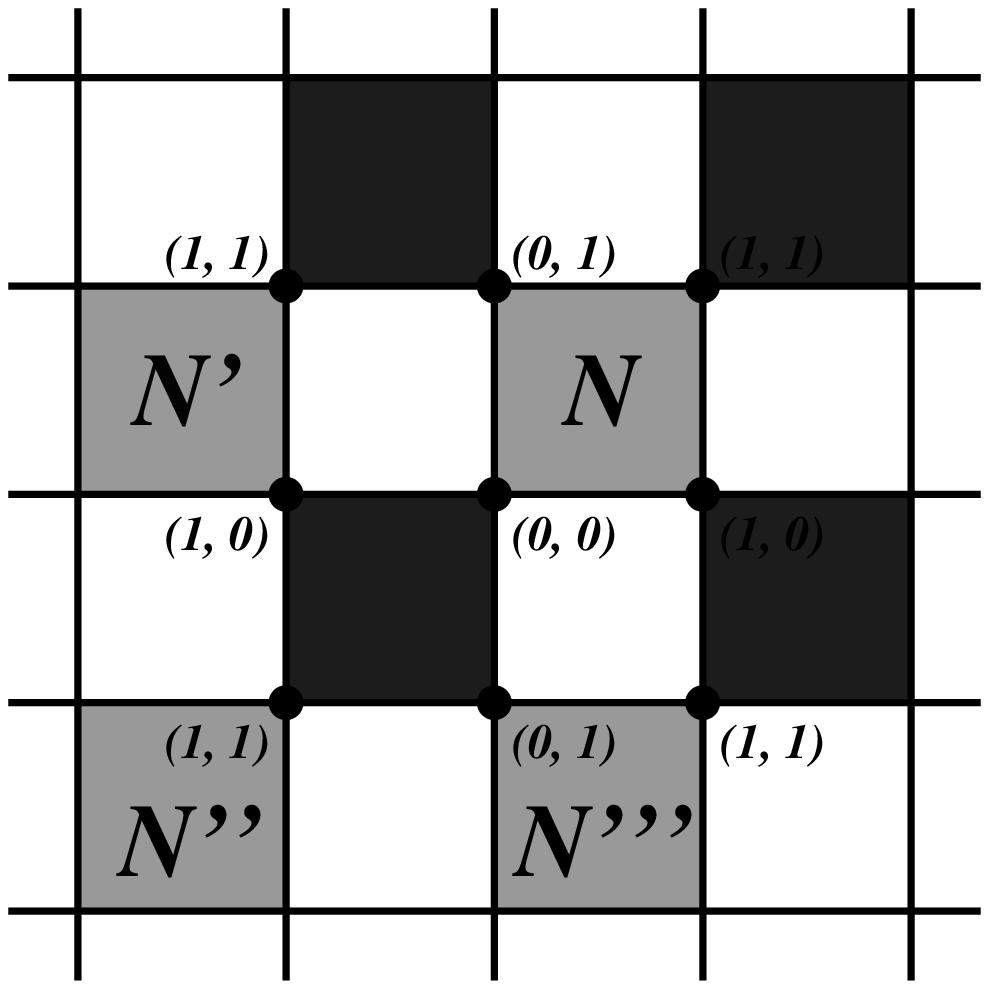} \\
\end{center}
\begin{small}
Fig.~1. Ichimatsu-patterned lattice in two dimensions.
\end{small}

We define a 'fundamental' spinor on a cell,

\begin{equation}
\psi_r(N) \equiv \xi_{r+2N} , ~~~  \bar{\psi}_r(N) \equiv \bar{\xi}_{r+2N}   . 
\end{equation}

The staggered Dirac operator is written as 

\begin{eqnarray}
D_{\rm st}(n,n')& =& \sum_{\mu} \eta_{\mu}(n)\frac{\delta_{n,n'+\hat{\mu}}U_{n,\mu}
-\delta_{n,n'-\hat{\mu}}U^{\dagger}_{n,\mu} }{2a} \nonumber  \\
  &= & \sum_{\mu,\vec{\varepsilon}}
(\Gamma_{\mu,\vec{\varepsilon}})_{r,r'} (d_{\mu,\vec{\varepsilon}})_{N,N'} ,
\end{eqnarray}
where $d_{\mu,\vec{\varepsilon}}$  is a generalized difference operator defined as

\begin{equation}
d_{\mu,\vec{\varepsilon}} \equiv \frac{1}{2^D}\sum_{\vec{\sigma}}(-1)^{\vec{\varepsilon}\cdot
\vec{\sigma}}\nabla_{\mu}^{\vec{\sigma}}  ,
\end{equation}
and $\nabla_{\mu}^{\vec{\sigma}}$  is a backward difference operator 
along $\mu$-direction at the place, $2N+\vec{\sigma}$. 
The coefficient matrices in eq. (5), 
\begin{eqnarray}
&(\Gamma_{\mu,\vec{\varepsilon}})
 \equiv((\sigma_3{}^{\varepsilon_1}
\otimes\cdots\otimes\sigma_3{}^{\varepsilon_D}) \label{generic Gamma} \nonumber\\
&\times(\sigma_3\otimes\cdots\otimes\sigma_3\otimes
\stackrel{\raise5pt\hbox{\scriptsize$\mu$}\mbox{ }}{\sigma_1}
\otimes{\bf 1}\otimes\cdots\otimes{\bf 1}))  , 
\end{eqnarray}
are base of SO(2D) Clifford algebra,
$
\gamma_{\mu} = \Gamma_{\mu,\vec{0}}
$
and 
$
\tilde{\gamma}_{\mu} = -i \Gamma_{\mu,\vec{e}_{\mu}}
$, and their odd  products.  Note that $\vec{\varepsilon}$ and $\vec{\sigma}$ 
are bit-valued D-dimensional vectors whose components are only 0 or 1. 
$\vec{e}_{\mu}$ is a unit vector 
along $\mu$-direction.

\section{SYMMETRIES IN A STAGGERED FERMION}

\subsection{Chiral symmetry}
For the staggered fermion, we can consider the exact chiral symmetry,
$$
\psi' = \exp({{\rm i}\theta \Gamma_{2D+1}}) \psi, ~~\bar{\psi}' = 
\bar{\psi} \exp({{\rm i}\theta \Gamma_{2D+1}}) .
$$
Under the transformation, it can be shown that 
 our Dirac action is invariant by the following relation,
$$
\{ \Gamma_{2D+1}, \Gamma_{\mu,\vec{\varepsilon}} \} = 0 , 
$$
where 
\begin{eqnarray}
(\Gamma_{2D+1})_{(r,r')} & \equiv &  (-i)^D 
(\gamma_1 \tilde{\gamma}_1 \cdots \gamma_D \tilde{\gamma}_D)_{(r,r')}  \nonumber \\
& = & (-1)^{|r|}\delta_{r,r'}.
\end{eqnarray}
This is equivalent to a  well-known even-odd transformation, 
\begin{equation}
\xi_n' = (-1)^{|n|},~~~~ \bar{\xi}_n' = -(-1)^{|n|}\bar{\xi}_n     ,
\end{equation}
and explains why there exists a mass-protect symmetry in a staggered fermion 
on an odd-dimensional lattice because  $2D$ is always even.

\subsection{Rotation} 
For cubic lattice, we can rotate  by the angle $\pi/2$ 
with respect to $\mu\nu$-plane as
\begin{eqnarray}
{\it R}_{\mu\nu}[\psi(N)] & = & a_{\mu\nu} \psi({R_{\mu\nu}[N])}, \nonumber \\
R_{\mu\nu}[\bar{\psi}(N)] & = & \bar{\psi}({R_{\mu\nu}[N])} a_{\mu\nu}^{\dagger},
\end{eqnarray}
where the prefactor of $\psi$, 
\begin{equation}
a_{\mu\nu} \equiv \frac{i}{2}\Gamma_{2D+1}(\tilde{\gamma}_{\mu} - \tilde{\gamma}_{\nu} )
(1+\gamma_{\mu}\gamma_{\nu}) ,
\end{equation}
is determined by the invariance of our Dirac action and the discrete rotation for a vector is  
\begin{eqnarray}
(R_{\mu\nu} [N])_{\rho} =  \left\{\begin{array}{ll}
           N_{\rho}  &   \rho \ne \mu,\nu \\
           -   N_{\nu}  &   \rho = \mu \\
             N_{\mu}  &   \rho = \nu    . 
              \end{array}\right.
\end{eqnarray}
The character   of this rotation is that $\gamma_{\mu}$ is transformed   
as a vector, 
\begin{eqnarray}
a_{\mu\nu} \gamma_{\rho}a_{\mu\nu}^{\dagger}= 
 \left\{\begin{array}{ll}
           \gamma_{\rho},  &   \rho \ne \mu,\nu \\
             - \gamma_{\nu},  &   \rho = \mu \\
             \gamma_{\mu},  &   \rho = \nu, 
              \end{array}\right.
\end{eqnarray}
but $\tilde{\gamma}_{\mu}$ is done under permutation, 
\begin{eqnarray}
a_{\mu\nu} \tilde{\gamma}_{\rho}a_{\mu\nu}^{\dagger}= 
 \left\{\begin{array}{ll}
           \tilde{\gamma}_{\rho},  &   \rho \ne \mu,\nu \\
             \tilde{\gamma}_{\nu},  &   \rho = \mu \\
             \tilde{\gamma}_{\mu},  &   \rho = \nu. 
              \end{array}\right.
\end{eqnarray}

\subsection{Parity}
For the odd-dimensional space, a usual  parity transformation does not imply 
disconnected one. Instead, to discuss general dimensions, we adopt $\mu$-directional    
reflection,  
\begin{eqnarray}
{\it P}_{\mu}[\psi(N)]  &=& \Gamma_{2D+1}\gamma_{\mu} \psi(P_{\mu}[N]), \nonumber \\ 
P_{\mu}[\bar{\psi}(N)]  & = &   \bar{\psi}(P_{\mu}[N]) \gamma_{\mu}\Gamma_{2D+1} , 
\end{eqnarray}
where 
\begin{eqnarray}
(P_{\mu} [N])_{\rho} = 
 \left\{\begin{array}{ll}
           N_{\rho}  &   \rho \ne \mu \\
           -   N_{\mu}  &   \rho = \mu  .
              \end{array}\right.
\end{eqnarray}
It is noted that under the transformation, (14), our action is invariant. 

\subsection{Scalar and pseudo-scalar modes}
Local meson operators are defined as 
\begin{equation}
\bar{\psi}_r(N) \Gamma_{r,r'} \psi_{r'}(N)  .
\end{equation}
We concentrate on scalar and pseudo-scalar modes. As discussion of sect. 3.2 and 3.3,  
only two scalar meson operators, 
\begin{equation}
M_1  = \bar{\psi}_r(N) \delta_{r,r'} \psi_{r'}(N)  ,
\end{equation}
and 
\begin{equation}
M_2 = \bar{\psi}_r(N) (\sum_{\mu}^D \tilde{\gamma}_{\mu})_{r,r'} \psi_{r'}(N)  ,
\end{equation}
are permitted. 
For pseudo-scalar meson operators, we find only the following two modes,  
\begin{equation}
M_3 = \bar{\psi}_r(N)  (\gamma_1\gamma_2\cdots\gamma_D)_{r,r'}  \psi_{r'}(N)  ,
\end{equation}
and
\begin{equation}
M_4 = \bar{\psi}_r(N) (\gamma_1\gamma_2\cdots\gamma_D  
\sum_{\mu}^D \tilde{\gamma}_{\mu})_{r,r'} \psi_{r'}(N) .
\end{equation}

\section{DISCUSSIONS}
On a D-dimensional cubic lattice, a naive discretized fermion has $2^D-1$ doublers  
per one mode.  The number of sites in a cell just corresponds to that of
 an original mode plus doublers which is exactly dimension of irreducible representation of 
SO(2D) Clifford algebra.  It leads to uniqueness of rotational, chiral and  
parity transformations.

For our staggered Dirac spinor, we should formulate  the fermion 
 gauge-covariantly.   It is formally possible if one connects the site of the fermion 
with the origin of the cell by a product of some link variables.  

Although we find only symmetric charge conjugation matrices, $C$, 
we can show the invariance under C, PT, CPT. 
That means that our staggered fermion is a vector theory.  

By using our formulation, we find an  scalar mode and a pseudo-scalar one  
in addition to well-known scalar and pseudo-scalar modes.  
With a bare mass,  operators are mixed  between two scalar modes
 and between pseudo-scalar ones.  To get the lowest meson mass,  
the mixing effect should be solved before taking  the continuum limit.  
In a cell, we may write down  
scalar modes by link variables.  
Our first clue to supersymmetric extention is the yukawa coupling term. 
From the  discussion in sect. 3 and the charge conjugation matrix,  
we have only one mode for the scalar in our action with nearest neighbor interactions.  
Exceptionally, for D=2 case, there are two candidates.   Further investigation is expected.

\end{document}